\documentclass[aps,twocolumn]{revtex4}
\usepackage{graphics}
\begin{document}
\newcommand{\bstfile}{aps} 
\newcommand{\bibs}{c:/PCTexv4/TexBiB/final}
\draft
\title{Studies on optimizing potential energy functions for maximal intrinsic hyperpolarizability}
\author{Juefei Zhou, Urszula B. Szafruga, David S. Watkins* and Mark G. Kuzyk}
\address{Department of Physics and Astronomy, Washington State University, Pullman, Washington  99164-2814; and *Department of Mathematics}

\begin{abstract}
We use numerical optimization to study the properties of (1) the class of one-dimensional potential energy functions and (2) systems of point charges in two-dimensions that yield the largest hyperpolarizabilities, which we find to be within 30\% of the fundamental limit.  We investigate the character of the potential energy functions and resulting wavefunctions and find that a broad range of potentials yield the same intrinsic hyperpolarizability ceiling of 0.709.
\end{abstract}

\maketitle

\section{Introduction}

Materials with large nonlinear-optical susceptibilities are central for optical applications such as telecommunications,\cite{wang04.01} three-dimensional nano-photolithography,\cite{cumps99.01,kawat01.01} and making new materials\cite{karot04.01} for novel cancer therapies.\cite{roy03.01}  The fact that quantum calculations show that there is a limit to the nonlinear-optical response\cite{kuzyk00.02,kuzyk01.01,kuzyk00.01,kuzyk03.01,kuzyk03.02,kuzyk04.02} is both interesting from the basic science perspective; and, provides a target for making optimized materials.  In this work, we focus on the second-order susceptibility and the underlying molecular hyperpolarizability, which is the basis of electro-optic switches and frequency doublers.

The fundamental limit of the off-resonance hyperpolarizability is given by,\cite{kuzyk00.01}
\begin{equation}\label{limit}
\beta_{MAX} = \sqrt[4]{3} \left( \frac {e \hbar} {\sqrt{m}}
\right)^3 \cdot \frac {N^{3/2}} {E_{10}^{7/2}} ,
\end{equation}
where $N$ is the number of electrons and $E_{10}$ the energy difference between the
first excited state and the ground state, $E_{10} = E_1 - E_0$.  Using Equation \ref{limit}, we can define the off-resonant intrinsic hyperpolarizability, $\beta_{int}$, as the ratio of the actual hyperpolarizability (measured or calculated), $\beta$, to the fundamental limit,
\begin{equation}\label{intrinsic-beta}
\beta_{int} = \beta / \beta_{MAX} .
\end{equation}
We note that since the dispersion of the fundamental limit of $\beta$ is also known, \cite{kuzyk06.03} it is  possible to calculate the intrinsic hyperpolarizability at any wavelength.  In the present work, we treat only the zero-frequency limit.

Until recently, the largest nonlinear susceptibilities of the best molecules fell short of the
fundamental limit by a factor of $10^{3/2}$, \cite{kuzyk03.02,Kuzyk03.05,Tripa04.01} so the very best molecules had a value of $\beta_{int} = 0.03$.  Since a Sum-Over-States (SOS) calculation of the hyperpolarizability\cite{orr71.01} using the analytical
wavefunctions of the clipped harmonic oscillator yields a value $\beta_{int} = 0.57$,\cite{Tripa04.01} the factor-of-thirty gap is not of a fundamental nature.  Indeed, recently, it was reported that a new molecule with asymmetric conjugation modulation has a measured value of $\beta_{int} = 0.048$.\cite{perez07.01}

To investigate how one might make molecules with a larger intrinsic hyperpolarizability, Zhou and coworkers used a numerical optimization process where a trial potential energy function is entered as an input, and the code iteratively deforms the potential energy function until the intrinsic hyperpolarizability, calculated from the resulting wavefunctions, converges to a local maximum.\cite{zhou06.01} In this work, a hyperbolic tangent function was used as the starting potential due to the fact that it is both asymmetric yet relatively flat away from the origin. This calculation was one-dimensional and included only one electron, so electron correlation effects were ignored.  Furthermore, the intrinsic hyperpolarizability was calculated using the new dipole-free sum-over-states expression\cite{kuzyk05.02} and only 15 excited states were included.  The resulting optimized potential energy function showed strong oscillations, which served to separate the spatial overlap between the energy eigenfunctions.  This led Zhou and coworkers to propose that modulated conjugation in the bridge between donor and acceptor ends of such molecules may be a new paradigm for making molecules with higher intrinsic hyperpolarizability.\cite{zhou06.01}

Based on this paradigm, P\'{e}rez Moreno reported measurements of a class of chromophores with varying degree of modulated conjugation.\cite{perez07.01}  The best measured intrinsic hyperpolarizability was $\beta_{int} = 0.048$, about 50\% larger than the best previously-reported.  Given the modest degree of conjugation modulation for this molecule, this new paradigm shows promise for further improvements.

In the present work, we extend Zhou's calculations to a larger set of starting potentials. To circumvent truncation problems associated with sum-over-states calculations, we instead determine the hyperpolarizability using a finite difference technique.  The optimization procedure is then applied to this non-perturbative hyperpolarizability.

To study the effects of geometry on the hyperpolarizability, Kuzyk and Watkins calculated the hyperpolarizability of various arrangements of point charges, representing nuclei, in two-dimensions using a two-dimensional Coulomb potential.\cite{kuzyk06.02}  In the present contribution, we apply our numerical optimization technique to determine the arrangement and charges of the nuclei in a planar molecule that maximize the intrinsic hyperpolarizability.

\section{Theory}

In our previous work, we used a finite-state SOS model of the hyperpolarizability that derives from perturbation theory (we used both the standard Orr and Ward SOS expression, $\beta_{SOS}$,\cite{orr71.01} and the newer dipole free expression, $\beta_{DF}$\cite{kuzyk05.02}).  The use of a finite number of states in lieu of the full infinite sums can result in inaccuracies, so, in the present work,  we use the non-perturbative approach, as follows.  We begin by solving the 1-d Schrodinger Equation on the interval $a<x<b$ for the ground state wavefunction $\psi(x,E)$ of an electron in a potential well defined by $V(x)$ and in the presence of a static electric field, $E$, that adds to the potential $\delta V = -exE$.  From this, the off-resonant hyperpolarizability is calculated with numerical differentiation, i.e. using finite differences, yielding
\begin{equation}\label{finite-diff-beta}
\beta_{NP} = \frac {1} {2} \left. \frac {\partial^2 \left( - \int_a^b \left| \psi(x,E) \right|^2 e x \, dx \right)}  {\partial E^2} \right|_{E=0}.
\end{equation}
Equation \ref{finite-diff-beta} is evaluated using the standard second-order approximation to the second
derivative:  
$$f''(z) \approx \frac{f(z+h) - 2f(z) + f(z-h)}{h^{2}}$$
with several $h$ values $h_{0}$, $h_{0}/5$, $h_{0}/25$, \ldots.  We
then refine these values by Richardson extrapolation \cite{kinca02.01}
and obtain our estimate from the two closest extrapolated values.

Our computational mesh consists of 200 quadratic finite elements with
a total of 399 degrees of freedom.  The potential energy function is a
cubic spline with 40 degrees of freedom.  Thus the numerical
calculations in regions where the potential function is represented by
3 points in the spline are covered by 15 elements with a total of
about 30 degrees of freedom.

Calculating $\beta_{int}$ from Equations \ref{finite-diff-beta}, \ref{intrinsic-beta} and \ref{limit} for a specific potential, we use an optimization algorithm that continuously
varies the potential in a way that maximizes $\beta_{int}$.  We also compute the matrix\cite{zhou06.01,kuzyk06.01}
\begin{equation}
\tau_{mp}^{(N)} = \delta_{m,p} - \frac {1} {2} \sum_{n=0}^{N} \left( \frac {E_{nm}} {E_{10}} + \frac {E_{np}} {E_{10}}\right) \frac {x_{mn}} {x_{10}^{max}} \cdot \frac {x_{np}} {x_{10}^{max}} , \label{tau}
\end{equation}
where $x_{10}^{max}$ is the magnitude of the fundamental limit of the position matrix element $x_{10}$ for a one electron system, and is given by,
\begin{equation}
x_{10}^{max} = \frac {\hbar} {\sqrt{2m E_{10}}} . \label{x10MAX}
\end{equation}
Each matrix element of $\tau^{(N)}$, indexed by $m$ and $p$, is a measure of how well the $(m,p)$ sum rule is obeyed when truncated to $N$ states.  If the sum rules are exactly obeyed, $\tau_{mp}^{(\infty)}=0$ for all $m$ and $p$.  We note that if the sum rules are truncated to an N-state model, the sum rules indexed by a large value of $m$ or $p$ (i.e. $m,p \sim N$) are disobeyed even when the position matrix elements and energies are exact.  We have found that the values of $\tau_{mp}^{(N)}$ are small for exact wavefunctions when $m<N/2$ and $p<N/2$.  So, when evaluating the $\tau$ matrix to test our calculations, we consider only the components $\tau_{m\leq N/2,p\leq N/2}^{(N)}$.

We observe that when using more than about 40 states in SOS calculations of the hyperpolarizability only a marginal increase of accuracy results when the potential energy function is parameterized with 400 degrees of freedom.  So, to ensure overkill, we use 80 states when calculating the $\tau$ matrix or the hyperpolarizability with an SOS expression so that truncation errors are kept to a minimum.  Since the hyperpolarizability depends critically on the transition dipole moment from the ground state to the excited states, we use the value of $\tau_{00}^{(40)}$ as one important test of the accuracy of the calculated wavefunctions.  Additionally, we use the standard deviation of $\tau^{(N)}$,
\begin{equation}
\Delta \tau^{(N)} = \frac{\sqrt{ \sum_{m=0}^{N/2} \sum_{p=0}^{N/2} \left( \tau_{mp}^{(N)} \right)^2 }} {N/2} , \label{Dtau}
\end{equation}
which quantifies, on average, how well the sum rules are obeyed in aggregate, making $\Delta \tau^{(N)}$ a broader test of the accuracy of a large set of wavefunctions.

Our code is written in MATLAB.  For each trial potential we use a quadratic
finite element method \cite{zienk05.01} to approximate the
Schr\"{o}dinger eigenvalue problem and the implicitly restarted
Arnoldi method \cite{soren92.01} to compute the wave functions and
energy levels.  To optimize $\beta$ we use the Nelder-Mead simplex
algorithm \cite{lagar98.01}.

As described in our previous work,\cite{zhou06.01} we perform optimization, but this time using the exact intrinsic hyperpolarizability $\beta = \beta_{NP}/\beta_{MAX}$, where $\beta_{MAX}$ is the fundamental limit of the hyperpolarizability, which is proportional to $E_{10}^{7/2}$.  To determine $E_{10} \equiv E_1 - E_0$, we also calculate the first excited state energy $E_1$.  

\section{Results and Discussions}

\begin{figure}
\includegraphics{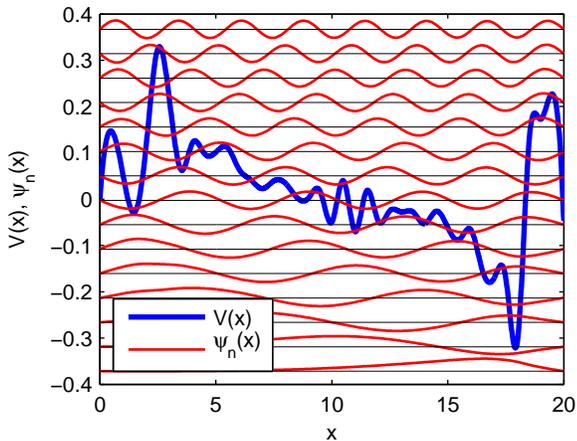}
\caption{Optimized potential energy function and first 15 wavefunctions after 7,000 iterations.  Starting potential is $V(x)=0$, using the non-perturbative hyperpolarizability for optimization.}
\label{fig:V=0-Potential-wavefuncitons}
\end{figure}
Figure \ref{fig:V=0-Potential-wavefuncitons} shows an example of the optimized potential energy function after 7,000 iterations when starting with the potential $V(x) = 0$ and optimizing the non-perturbative intrinsic hyperpolarizability $\beta_{NP}/ \beta_{MAX}$ as calculated with Equation \ref{finite-diff-beta}.  Also shown are the eigenfunctions of the first 15 states computed from the optimized potential.  First, we note that the potential energy function shows the same kinds of wiggles as in our original paper,\cite{zhou06.01} though not of sufficient amplitude to localize the wavefunctions.

For the starting potentials we have investigated, our results fall into two broad classes.  In the first, three common features are: (1) The best intrinsic hyperpolarizabilities are near $\beta_{int} = 0.71$; (2) the best potentials have a series of wiggles; and (3) the systems behave as a limited-state model.  In the second class of starting potentials, (2) the wiggles are much less pronounced and (3) more states contribute evenly.  Figure \ref{fig:V=0-Potential-wavefuncitons} is an example of a Class II potential.  However, in both classes, the maximum calculated intrinsic hyperpolarizability appears to be bounded by $\beta_{int} = 0.71$.   Using the set of potentials from both classes that lead to optimized $\beta_{NP}/\beta_{MAX}$, we calculate the lowest 80 eigenfunctions and eigenvalues, from which we calculate $\beta_{DF}$ and $\beta_{SOS}$.  In most cases, we find that the three different formulas for $\beta$ converge to the same value when only the first 20 excited states are used (using 80 states, the three are often the same to within at least 4 decimal places) and $\tau_{00} \approx 10^{-4}$, showing that the ground state sum rules are well obeyed.  Furthermore, the rms deviation of the $\tau$ matrix when including 40 states leads to $\tau^{(80)}<0.001$.

\begin{figure}
\includegraphics{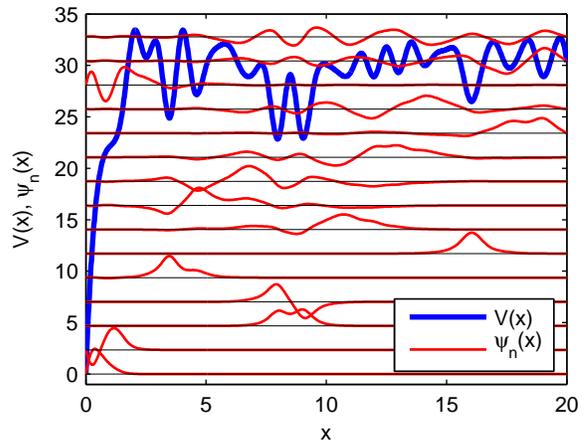}
\caption{Optimized potential energy function and first 15 wavefunctions after 8,000 iterations.  Starting potential is $V(x)=\tanh(x)$, using the non-perturbative hyperpolarizability for optimization.}
\label{fig:vtanh40}
\end{figure}
Figure \ref{fig:vtanh40} shows an example of the optimized potential energy function when starting with the potential $V(x) = \tanh x$ and optimizing the exact (non-perturbative) intrinsic hyperpolarizability.  Also shown are the eigenfunctions of the first 15 states computed with the optimized potential.  First, we note that the potential energy function shows the same kinds of wiggles as in our original paper;\cite{zhou06.01} and only 2 excited state wavefunctions and the ground state are localized in the first deep well - placing this system in Class I.

The observation that such potentials lead to hyperpolarizabilities that are near the fundamental limit motivated Zhou and coworkers to suggest that molecules with modulated conjugation may have enhanced intrinsic hyperpolarizabilities.\cite{zhou06.01} A molecule with a modulated conjugations bridge between the donor and acceptor end was later shown to have record-high intrinsic hyperpolarizability.\cite{perez07.01}  As such, this result warrants a more careful analysis.

\begin{figure}
\includegraphics{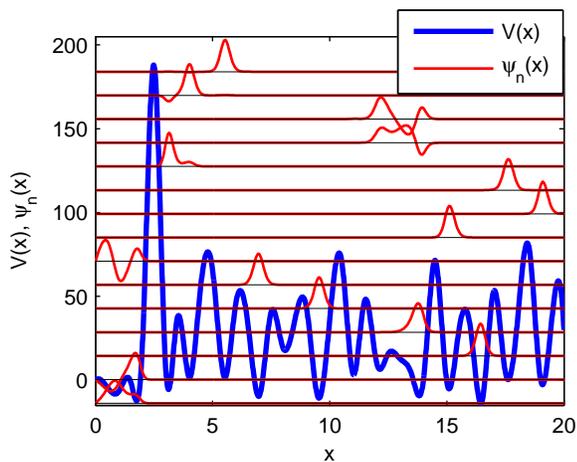}
\caption{Optimized potential energy function using $\beta_{DF}$ and first 15 wavefunctions after 7,000 iterations.  Starting potential is the $\tanh (x)$ potential.  The final potential (shown above) we refer to as the Zhou potential.}
\label{fig:juefei0}
\end{figure}

It is worthwhile to compare our present results characterized by Figure \ref{fig:vtanh40} with our past work,\cite{zhou06.01} particularly for the purpose of examining the impact of the approximations used in the previous work.\cite{zhou06.01} Figure \ref{fig:juefei0} shows the optimized potential and wavefunctions obtained by Zhou and coworkers using a 15-state model and optimizing the dipole-free intrinsic hyperpolarizability.  Since only 15 states were used, the SOS expression for $\beta$ did not fully converge; making the result inaccurate as suggested by the fact that $\beta_{SOS}$ and $\beta_{DF}$ did not agree.  However, since the code focused on optimizing the dipole-free form of $\beta$, and $\tau_{00}$ was small when $\beta_{int}$ was optimized, the dipole-free expression may have converged to a reasonably accurate value while the commonly-used SOS expression was inaccurate.  Indeed, it was found that $\beta_{DF} \approx 0.72$ -  in contrast to our more precise present calculations using the non-perturbative approach, which yields $\beta_{NP} < 0.71$.  So, the fact that our more precise calculations, which do not rely on a sum-over states expression, agree so well with the 15-state model suggests that in both cases, the limit for a one-dimensional single electron molecule is just over $\beta \approx 0.7$.  This brute force calculation serves as a numerical illustration of the observation that the limiting value of $\beta$ is the same for an exact non-perturbation calculation and for a calculation that truncates the SOS expression, which presumedly should lead to large inaccuracies.\cite{Champ05.01,kuzyk05.01} At minimum, this result supports the existence of fundamental limits of nonlinear susceptibilities that are in line with past calculations.

To state Zhou's approach more precisely,\cite{zhou06.01} the calculations optimized the very special case of the intrinsic hyperpolarizability for a 15 state model for a potential energy function that is parameterized with 20 spline points.  As such, the potential energy function can at most develop about 20 wiggles.  As a consequence, there are enough degrees of freedom in the potential energy function to force the 15 states to be spatially well separated.  Interestingly, after optimization, only two excited states overlap with ground state, allowing only these two states to have nonzero transition dipole moments with each other and the ground state -- forcing the system into a three-level SOS model for $\beta_{DF}$.  This behavior is interesting in light of the three-level ansatz, which asserts that only three states determine the nonlinear response of a system when it is near the fundamental limits.

It is interesting to compare the exact non-perturbation calculation, which does not depend on the excited state wavefunctions (Figure \ref{fig:vtanh40}) and Zhou's {\em contrived} system of 15 states (Figure \ref{fig:juefei0}).  Both cases have wiggles and the wavefunctions appear to be mostly non-overlapping.  So, for the first 15 states, the wavefunctions appear similarly localized.  The situation becomes more interesting when 80 states are included in calculating the hyperpolarizability for Zhou's potential or when the exact non-perturbative approach is used.  The first line in Table \ref{tab:zhouV} summarizes the results with Zhou's potential and 80 states.
\begin{table}\caption{Evolution of Zhou's Potential. $\beta_s$ is the hyperpolarizability of the starting potential using 80 states while the other ones are after optimization of $\beta_{NP}$.\label{tab:zhouV}}
\begin{tabular}{c c c c c c c}
  \hline
  Number of & $\beta_{S}$ & $\beta_{SOS}$ & $\beta_{DF}$ & $\beta_{NP}$ & $\tau_{00}^{(80)}$ & $\Delta \tau^{(80)}$ \\
 Iterations &  & & & & ($\times 10^{-5}$) & ($\times 10^{-4}$) \\
  \hline
   0 & 0.5612 & 0.5612 & 0.5607 & 0.5612 & 11.2  & 15 \\
  1000 & 0.5612 & 0.7087 & 0.6682 & 0.7083 & 1810 &  40\\
  \hline
\end{tabular}
\end{table}

First, let's focus on the sum-over-states results.  Clearly, when 80 states are used in the calculation, it is impossible for the excited state wavefunctions to not overlap with each other, so the three-level approximation to $\beta$ breaks down.  According to the three-level ansatz, we would expect the hyperpolarizability to get smaller.  Indeed, the additional excited states result in a smaller hyperpolarizability ($\approx 0.56$).  Note that the exact and SOS expressions agree with each other and that $\tau_{00}^{(80)}$ and  $\Delta \tau^{(80)}$ are small.

\begin{figure}
\includegraphics{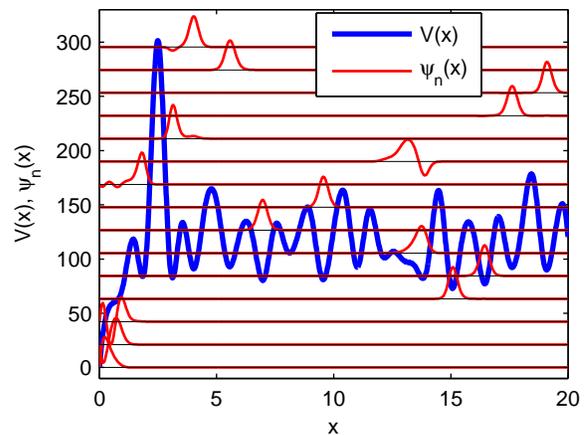}
\caption{Optimized potential energy function and first 15 wavefunctions after 1,000 iterations.  Starting potential is Zhou's potential, using the non-perturbative hyperpolarizability for optimization.}
\label{fig:juefei1000}
\end{figure}

Figure \ref{fig:juefei1000} shows the result after 1000 iterations, using Zhou's potential as the starting potential and using the non-perturbative hyperpolarizability for optimization.  First, the non-perturbative hyperpolarizability reaches just under 0.71, but, the SOS and dipole-free expressions do not agree with each other.  Furthermore, both convergence metrics ($\tau_{00}^{(80)}$ and $\Delta \tau^{(80)}$) are larger than before optimization.  It would appear that for Zhou's potential, even 80 states are not sufficient to characterize the nonlinear susceptibility when a sum-over-states expression is used (either dipole free or traditional SOS expression - though the SOS expression agrees better with the non-perturbative approach). 

Interestingly, the optimized potential energy function still retains the wiggles and the wave functions are still well separated.  This result is consistent with the suggestion of Zhou and coworkers that modulation of conjugation may be a good design strategy for making large-hyperpolarizability molecules.  We note that wiggles in the potential energy function are not required to get a large nonlinear-optical response; but, appears to be one way that Mother Nature optimizes the hyperpolarizability.  Since this idea has been used to identify molecules with experimentally measured record intrinsic hyperpolarizability,\cite{perez07.01} the concept of modulation of conjugation warrants further experimental studies.

\begin{figure}
\includegraphics{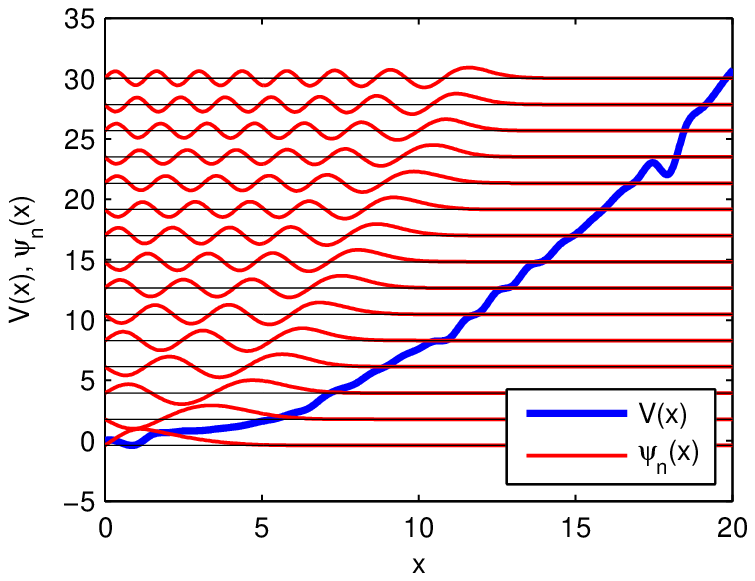}
\caption{Optimized potential energy function and first 15 wavefunctions after 8,000 iterations.  Starting potential is $V(x)=x^2$, using the non-perturbative hyperpolarizability for optimization.}
\label{fig:CHO8300}
\end{figure}

\begin{figure}
\includegraphics{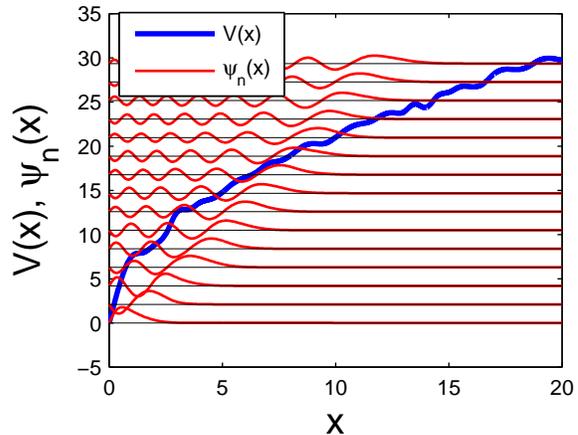}
\caption{Optimized potential energy function and first 15 wavefunctions after 8,000 iterations.  Starting potential is $V(x)=\sqrt{x}$, using the non-perturbative hyperpolarizability for optimization.}
\label{fig:sqrt8000}
\end{figure}

As a case in point that non-wiggly potentials can lead to a large nonlinear susceptibility is the clipped harmonic oscillator, which we calculated to have an intrinsic hyperpolarizability of about 0.57.\cite{Tripa04.01}  Figure \ref{fig:CHO8300} shows the optimized non-perturbative hyperpolarizability when using a clipped harmonic oscillator as the starting potential.  The properties of all of the optimized potentials are summarized in Table \ref{tab:Vsumary}.   The clipped square root function also has a large hyperpolarizability (0.69).  The optimized potential is shown in Figure \ref{fig:sqrt8000}.  In these cases, the amplitude of the wiggles are small and all the wavefunctions overlap.  So, these fall into Class II.  Note that the lack of wiggles shows that they are not an inevitable consequence of our numerical calculations.
\begin{table}\caption{Summary of calculations with different starting potentials.  $\beta_s$ is the hyperpolarizability of the starting potential while the other ones are after optimization.\label{tab:Vsumary}}
\begin{tabular}{c c c c c c c}
  \hline
  Function & $\beta_{S}$ & $\beta_{SOS}$ & $\beta_{DF}$ & $\beta_{NP}$ & $\tau_{00}^{(80)}$ & $\Delta \tau^{(80)}$ \\
 $V(x)$ &  & & & & ($\times 10^{-5}$) & ($\times 10^{-4}$) \\
  \hline
   0 & 0 & 0.7089 & 0.7089 & 0.7089 & $37.8 $ & $5.33 $ \\
  $30\tanh(x)$ & 0.67 & 0.7084 & 0.6918 & 0.7083 & 779 & 11.8 \\
  $x$ & 0.66 & 0.7088 & 0.7072 & 0.7088 & 78.7 & 8.79 \\
  $x^2$ & 0.57 & 0.7089 & 0.7085  & 0.7088 & 18.6 & 703 \\
  $x^{1/2}$ & 0.68 & 0.7087 & 0.7049 & 0.7087 & 190 & 9.76 \\
  $x + \sin(x)$ & 0.67 & 0.7088 & 0.7073 & 0.7088 & 75.0 & 8.46 \\
  $x + 10 \sin(x)$ & 0.04 & 0.7085 & 0.7085 & 0.7085 & 1.65 & 7.78 \\
  \hline
\end{tabular}
\end{table}

We may question whether small wiggles in the potential energy function lead to large amplitude wiggles as an artifact of our numerical optimization technique.  To test this hypothesis, we used the trial potential energy function $x+\sin(x)$, where the wiggle amplitude is not large enough to cause the wavefunctions to localize at the minima.  The optimized potential energy function retains an approximately linear from with only small fluctuation.  In fact, the results are very similar to what we found for the linear starting potential and the wiggles do not affect the final result.  The similarity between these cases can be seen in Table \ref{tab:Vsumary}.

\begin{figure}
\includegraphics{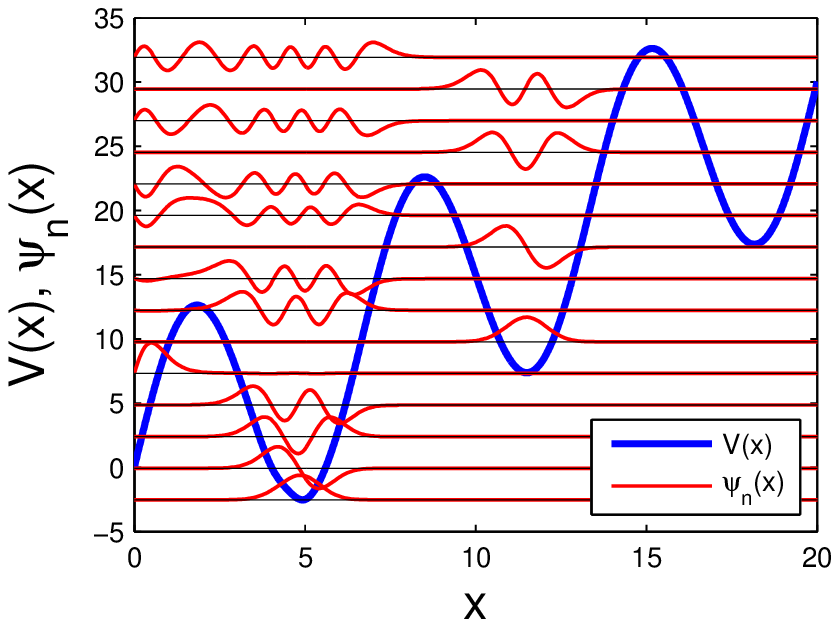}
\includegraphics{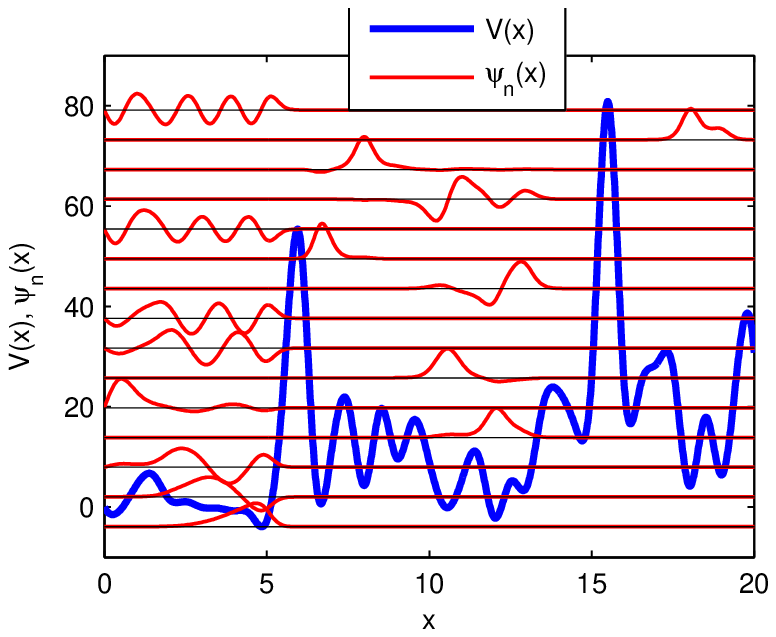}
\caption{Potential energy function and first 15 wavefunctions before (top) and after (bottom) 3,000 iterations.  Starting potential is of the form $V(x)=x + 10 \sin(x)$, using the non-perturbative hyperpolarizability for optimization.}
\label{fig:x10sinx3000}
\end{figure}

Next, we test a starting potential with large wiggles as shown in the upper portion of Figure \ref{fig:x10sinx3000}.  The lower energy eigenfunctions are found to be localized mostly in the first two wells.  In fact, the lowest four energy eigenfunctions are well approximated by harmonic oscillator wavefunctions, which are centrosymmetric.  As a result, the first excited state holds most of the oscillator strength and the value of the intrinsic hyperpolarizability is only 0.04. 

After 3000 interactions, this Class I potential energy function has high amplitude wiggles at a wavelength that is significantly shorter than the wavelength of the initial sine function (bottom portion of Figure \ref{fig:x10sinx3000}).  In common with the optimized $\tanh(x)$ function, the wiggles are of large but almost chaotically varying amplitude.  This leads to wavefunctions that are spatially separated.  While the wavefunctions are not as well separated as we find for the $\tanh(x)$ starting potential, the optimized potential yields only two dominant transition from the ground state; so, this system is well approximated by a three-level model.  As is apparent from Table \ref{tab:Vsumary}, the ground state sum rule (characterized by $\tau_{00}^{(80)}$) is better obeyed in this optimized potential than in any others.  So, the wavefunctions are accurate and all of the values of $\beta$ have converged to the same value, suggesting that this calculation may be the most accurate of the set

Our results bring up several interesting questions.  First, all of our extensive numerical calculations, independent of the starting potential, yield an optimized intrinsic hyperpolarizability with an upper bound of 0.71, which is about thirty percent lower than what the sum rules allow.  Since numerical optimization can settle in to a local maximum, it is possible that all of the starting potentials are far from the global maximum of $\beta_{int}=1$.  Indeed, since most potentials lead to systems that require more than three dominant states to express the hyperpolarizability, this may in itself be an indicator that we are not at the fundamental limit precisely because these systems have more than three states.  Indeed, the original results of Zhou and coworkers frames the problem in a way (i.e. a 15-level model in a potential limited to about 20 wiggles) that allows a solution to the optimization problem to lead to three dominant states.  So, while it may be argued that this system is contrived and unphysical, we have found value in trying such toy models when testing various hypotheses.  This toy model
\begin{itemize}
\item{leads to a three-level system as the three-level ansatz proposes}
\item{has the same qualitative properties as more precise methods}
\item{has given insights into making new molecules with record-breaking intrinsic hyperpolarizability}
\end{itemize}
Given the complexity of calculating nonlinear-susceptibilities, our semi-quantitative method may be a good way of generating new ideas.

The three-level ansatz proposes that at the fundamental limit, all transitions are negligible except between three dominant states.  There appears to be no proof of the ansatz aside from the fact that it leads to an accurate prediction of the upper bound of nonlinear susceptibilities, both calculated and measured.  To understand the motivation behind the ansatz, it is useful to understand how the two-level model optimizes the polarizability, $\alpha$, without the need to rely on any assumptions.  This is trivial to show by using the fact that the polarizability depends only on the positive-definite transition moments, $\left<0 \right| x \left| n \right> \left< n \right| x \left| 0 \right> $, the same parameters that are found in the ground state sum rules.\cite{kuzyk06.01}

For nonlinear susceptibilities, the situation is much more complicated because the SOS expression  depends on quantities such as $\left<0 \right| x \left| n \right> \left<n \right| x \left| m \right>\left< m \right| x \left| 0 \right> $, where these terms can be both positive and negative.  Furthermore, the sum rules that relate excited states moments to each other allow for these moments to be much larger than transition moments to the ground state.  So, it would seem plausible that one could design a system with many excited states in a way that all of the transition moments between excited states would add constructively to yield a larger hyperpolarizability than what we calculate with the three-level ansatz.  None of our numerical calculations, independent of the potential energy function, yield a value greater than 0.71.  Since our potential energy functions are general 1-dimensional potentials (i.e. the potentials are not limited to Coulomb potentials, nor are the wavefunctions approximated as is common in standard quantum chemical computations), our calculations most likely span a broader range of possible wavefunctions leading to a larger variety of states that contribute to the hyperpolarizability.

However, there appear to be local maxima associated with systems that behave as a three-level system and others with many states, and, the maximum values both are 0.71.  It is interesting that so may different sets of transition moments and energies can yield the exact same local maximum.  To gain a deeper appreciation of the underlying physics, let's consider the transition moments and energies in the sum-over-states expression for the hyperpolarizability as adjustable parameters.  For a system with $N$ states, there are $N-1$ energy parameters of the form $E_n - E_0$.  The moment matrix $x_{ij}$ has $N^2$ components.  If the matrix is real, there are $(N^2-N)/2$ unique off-diagonal terms and $N$ diagonal dipole moments.  Since all dipole moments appear as differences of the form $x_{nn}-x_{00}$, there are only $N-1$ dipole moment parameters.  Therefore, the dipole matrix is characterized by $(N^2-N)/2 + N - 1 = (N+2)(N-1)/2$ parameters.  Combining the energy and dipole matrix parameters, there are a total of $(N+2)(N-1)/2 + N-1$ parameters.

The N-state sum rules are of the form:
\begin{eqnarray}
& & \sum_{n=0}^\infty \left( E_n - \frac {1} {2} \left( E_m + E_p \right) \right)\left<m \right| x \left| n \right> \left<n \right| x \left| p \right> \\ \nonumber
& = & \frac {\hbar^2 N} {2m} \delta_{m,p}, 
\end{eqnarray}
so the sum rules comprise a total of $N^2$ equations (i.e. an equation for each $(m,p)$).  If the sum rules are truncated to $N$ states, the sum rule indexed by $(m=N,p=N)$ is nonsensical because it contradicts the other sum rules.  Furthermore, if the transition moments are real, then $x_{mp} = x_{pm}$, so only $(N^2 - N)/2$ of the equations are independent.  As such, there are a total of $(N^2 - N)/2 + N-1 = (N+2)(N-1)/2$ independent equations.

Since the SOS expression for the nonlinear-susceptibility has $(N+2)(N-1)/2 + N-1$ parameters and the sum rules provide $(N+2)(N-1)/2$ equations, the SOS expression can be reduced to a form with $N-1$ parameters.  For example, the three-level model for the hyperpolarizability, which is expressed in terms of 7 parameters, can be reduced to two parameters using 5 sum rule equations.  In practice, however, even fewer sum rule equations are usually available because some of them lead to physically unreasonable consequences.  While the $(N,N)$ sum rule is clearly unphysical due to truncation, sum rule equations that are near equation $(N,N)$ may also be unphysical.  In the case of the three-level model, it is found that the equation $(2,1)$ allows for an infinite hyperpolarizability, so that equation is ignored on the grounds that it violates the principle of physical soundness.\cite{kuzyk05.01,kuzyk06.01,kuzyk06.03}  This leads to a hyperpolarizability in terms of 3 variables, which are chosen to be $E_{10}$, $E = E_{10}/E_{20}$, and $X = x_{10}/x_{10}^{MAX}$.  The expression is then maximized with respect to the two parameters $E$ and $X$, leaving the final result a function of $E_{10}$.

We conclude that the SOS expression for the hyperpolarizability can be expressed in terms of at least $N-1$ parameters; so, it would appear that as more levels are included in the SOS expression, there are more free parameters that can be varied without violating the sum rules.  As $N \rightarrow \infty$, there are an infinite number of adjustable parameters.  So, it is indeed puzzling that the three-level ansatz yields a fundamental limit that is consistent with all of our calculations for a wide range of potentials, many of which have many excited states.  It may be that we are only considering a small subset of potential energy functions; or, perhaps the expression for the hyperpolarizability depends on the parameters in such a way that large matrix elements contribute to the hyperpolarizability with alternating signs so that the big terms cancel.  This is a puzzle that needs to be solved if we are to understand what makes $\beta$ large.

To investigate whether the limiting behavior is due to our use of 1-dimensional potentials, we have also optimized the intrinsic hyperpolarizability in two-dimensions.  In this case, we focus on the largest tensor component, $\beta_{xxx}$ and describe the potential as a superposition of point charges.  As described in the literature,\cite{kuzyk06.02} we solve the two-dimensional Schr\"{o}dinger eigenvalue problem,
\begin{equation}\label{eq:schroedinger}
-\frac{\hbar^{2}}{2m} \nabla^{2}\Psi + V\Psi = E \Psi ,
\end{equation}
for the lowest ten to 25 energy eigenstates, depending on the degree of convergence of the resulting intrinsic hyperpolarizability.  We use the two-dimensional logarithmic Coulomb potential, which for $k$ nuclei with charges $q_{1}e$, \ldots, $q_{k}e$ located at points $s^{(1)}$, \ldots, $s^{(k)}$, is given by
\begin{equation}\label{eq:2Dpotential}
V(s) = \frac{e^{2}}{L}\sum_{j=1}^{k}q_{j}\log \|s - s^{(j)} \|, 
\end{equation}
where $L$ is a characteristic length.  With $L = 2 \AA$, the force due to a charge at distance $2 \AA$ is the
same as it would be for a 3D Coulomb potential.

We discretize the eigenvalue problem given by Equation \ref{eq:schroedinger} using a
quadratic finite element method \cite{zienk05.01,atkin01.01} and solve the
resulting matrix eigenvalue problem for the ten to 25 smallest energy eigenvalues
and corresponding eigenvectors by the implicitly-restarted Arnoldi
method \cite{soren92.01} as implemented in ARPACK \cite{lehou98.01}.  
Each eigenvector yields a wave function $\Psi_{n}$ corresponding to
energy level $E_{n}$.  The moments 
$$x_{mn} = \int_{-\infty}^{\infty}\!\!\int_{-\infty}^{\infty} s_{1}
\Psi_{m}(s_{1},s_{2}) \Psi_{n}(s_{1},s_{2})\, ds_{1}ds_{2}$$
are computed, and these and the energy levels $E_{n}$ are used to
compute $\beta$ 

\begin{figure}
\includegraphics{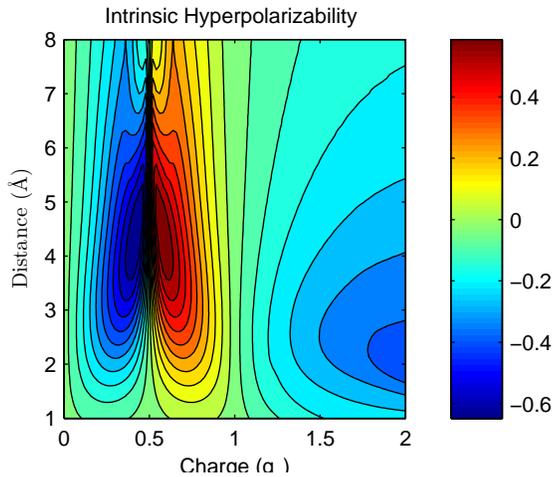}
\caption{The intrinsic hyperpolarizability of two nuclei as a function of the distance between them and the charge of one nucleus, $q_1$ where $q_1+q_2 = +e $.}
\label{fig:2chargeCountour}
\end{figure}

Figure \ref{fig:2chargeCountour} shows the intrinsic hyperpolarizability of a two-nucleus molecule plotted as a function of the distance between the two nuclei and nuclear charge $q_1$.  The total nuclear charge is $q_1 + q_2 = +e$, and is expressed in units of the proton charge, $e$.  Three extrema are observed.  The positive peak parameters are $\beta_{int} = 0.649$ for $q_1 = 0.58$ and $d = 4.36 \AA$.  The negative one yields $\beta_{int} = -0.649$ for $q_1 = 0.42$ and $d = 4.36 \AA$.  The local negative peak that extends past the graph on the right reaches its maximum magnitude of $\beta_{int} = -0.405$ at $q_1 = 2.959$ and $d = 2.0 \AA$. 

Applying numerical optimization to the intrinsic hyperpolarizability using the charges and separation between the nuclei as parameters, we get $\beta_{int} = 0.654$ at $d = 4.539 \AA$, and $q_1 = 0.430$ when the starting parameters are near the positive peak; and $\beta_{int} = -0.651$, $d = 4.443 \AA$, and $q_1 = 0.572$ when optimization gives the negative peak.  The peak parameters are the same within roundoff errors when optimization or plotting is used, confirming that the optimization procedure yields the correct local extrema.

\begin{figure}
\includegraphics{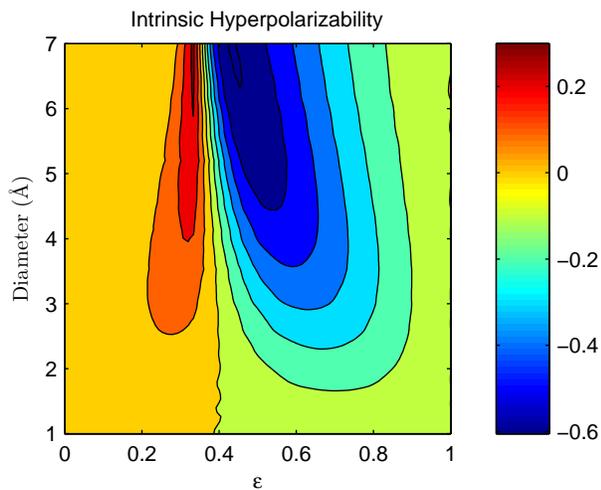}
\caption{The intrinsic hyperpolarizability of three evenly-spaced nuclei on a circle as a function of the circle's diameter and the charge, $\epsilon$ (in units of $e$), on one of the nuclei.  The charge on each of the other nuclei is $e (1 - \epsilon)/2$.}
\label{fig:3chargeContour}
\end{figure}

Figure \ref{fig:3chargeContour} shows the intrinsic hyperpolarizability of an octupolar-like molecule made of three evenly-spaced nuclei on a circle plotted as a function of the circle's diameter and charge fraction $\epsilon$ ($q = \epsilon e$).  The charge on each of the other nuclei is $e (1 - \epsilon)/2$.
The positive peak at $\epsilon = 0.333$ and diameter $D = 6.9 \AA$ has a hyperpolarizability $\beta_{int} = 0.326$, while $\beta_{int} = -0.605$ for a charge fraction $\epsilon = 0.44$ and a diameter $D = 6.8 \AA$.

When the positions and magnitudes of the three charges are allowed to move freely, the best intrinsic hyperpolarizability obtained using numerical optimization is $\beta_{int} = 0.685$ for charges located at $\vec{r}_1 = (0,0)$. $\vec{r}_2 = (-4.87 \AA,0.33\AA)$, and $\vec{r}_3 = (-9.57\AA,-0.16\AA)$; with charges $q_1 = 0.43e$, $q_2 = 0.217e$, and $q_3=0.351e$.  There are only small differences in the optimized values of $\beta_{int}$ depending on the starting positions and charges; and the best results are for a ``molecule" that is nearly linear along the x-direction.  This is not surprising given that the $xxx$-component of $\beta_{int}$ is the optimized quantity.

The two-dimensional analysis illustrates that numerical optimization correctly identifies the local maxima (peaks and valleys) and that the magnitude of maximum intrinsic hyperpolarizability (0.65 vs 0.68) is close to the maximum we get for the one-dimensional optimization of the potential energy function (0.71).  All computations we have tried, including varying the potential energy function in one dimension or moving around point charges in a plane all yield an intrinsic hyperpolarizability that is less than $0.71$.

An open question is the origin of the factor-of-thirty gap between the best molecules and the fundamental limit, which had remained firm for decades through the year 2006.  Several of the common proposed explanations, such as vibronic dilution, have been eliminated.\cite{Tripa04.01} Perhaps it is not possible to make large-enough variations of the potential energy function without making the molecule unstable.  Or, perhaps there are subtle issues with electron correlation, which prevents electrons from responding to light with their full potential.  The fact that the idea of modulation of conjugation has lead to a 50\% increase over the long-standing ceiling - reducing the gap to a factor of twenty - makes it a promising approach for potential further improvements.  Continued theoretical scrutiny, coupled with experiment, will be required to confirm the validity of our approach.

\section{Conclusions}

There appear to be many potential energy functions that lead to an intrinsic hyperpolarizability that is near the fundamental limit.  These separate into two broad classes: one in which wiggles in the potential energy function forces the eigenfunctions to be spatially separated and a second class of monotonically varying wavefunctions with small or no wiggles that allow for many strongly overlapping wavefunctions.  Interestingly, all these one-dimensional ``molecules" have the same maximal intrinsic hyperpolarizability of $0.71$.  It is puzzling that the three-level ansatz correctly predicts the fundamental limit even when the ansatz does not apply.  A second open question pertains to the origin of the long-standing factor of 30 gap between the fundamental limit and the best molecules.  The idea of conjugation modulation may be one promising approach for making wiggly potential energy profiles that lead to molecules that fall into the gap.  Given that there are so many choices of potential energy functions that lead to maximal intrinsic hyperpolarizability, it may be possible to engineer many new classes of exotic molecules with record intrinsic hyperpolarizability.  

{\bf Acknowledgements: } MGK thanks the National Science Foundation (ECS-0354736) and Wright Paterson Air Force Base for generously supporting this work.


\clearpage

\end{document}